\documentclass[baaa]{baaa}

\usepackage[pdftex]{hyperref}
\usepackage{subfigure}
\usepackage{natbib}
\usepackage{helvet,soul}
\usepackage[font=small]{caption}

\contriblanguage{1}

\contribtype{1}

\thematicarea{7}

\received{\ldots}
\accepted{\ldots}

\title{Kinematic patterns of the enriched gas phase \\ in the Local Group {\sc Hestia} simulations}

\titlerunning{Kinematics of the enriched gas phase in the {\sc Hestia} simulations}

\author{
L. Biaus\inst{1},
S.E. Nuza\inst{2,3},
C. Scannapieco\inst{1,3},
F.G. Iza\inst{1,2}
\&
E. Lozano\inst{1}
}

\authorrunning{Biaus et al.}

\contact{lbiaus@df.uba.ar}

\institute{
Departamento de Física, Facultad de Ciencias Exactas y Naturales, UBA, Argentina
\and
Instituto de Astronom{\'\i}a y F{\'\i}sica del Espacio, CONICET--UBA, Argentina
\and
Consejo Nacional de Investigaciones Cient\'ificas y T\'ecnicas, Argentina
}

\resumen{Las observaciones de {\em absorbers} en el Grupo Local sugieren la existencia de un dipolo de velocidades en la dirección general del baricentro-antibaricentro, lo que puede interpretarse como evidencia de un flujo general de material hacia el centro de masa del grupo. En este trabajo, estudiamos la cinemática del gas en el Grupo Local utilizando una de las realizaciones de alta resolución de las simulaciones {\sc Hestia}, con un enfoque en la evidencia proporcionada por diferentes especies iónicas. Nuestra simulación incluye la cosmografía correcta para una región similar al Grupo Local y una velocidad radial relativa entre las candidatas a la  Vía Láctea y Andrómeda consistente con la observada. Examinamos la distribución y la cinemática de seis especies iónicas (H\,{\sc i}, C\,{\sc iv}, Si\,{\sc iii}, O\,{\sc vi}, O\,{\sc vii} y O\,{\sc viii}) y sus huellas en mapas sintéticos del cielo construidos desde los marcos de referencia más utilizados por los observadores. Nuestros resultados indican la presencia de dicho dipolo para el gas fuera del halo de la Vía Láctea, favoreciendo un paradigma en el que la Vía Láctea se mueve contra el gas en dirección al baricentro, mientras que se aleja del gas en la dirección opuesta. Este patrón es más claro para los iones más altos de oxígeno, que trazan gas caliente. Por otro lado, observamos una ligera asimetría en los perfiles de presión en ambas direcciones, indicando presiones más altas en las regiones internas del Grupo Local.}

\abstract{
Observations of intergalactic absorbers in the Local Group suggest the existence of a velocity dipole in the general barycentre--antibarycentre direction which can be interpreted as evidence of a general flow of material towards the group's centre of mass. In this work, we study the kinematics of gas in the Local Group using one of the high-resolution realisations of the {\sc Hestia} simulations with a particular focus on the evidence left by different ionic species. Our simulation includes the correct cosmography for a region similar to the Local Group and a relative radial velocity between the candidate Milky Way and Andromeda galaxies consistent with the observed one. We examine the distribution and kinematics of six ionic species (H\,{\sc i}, C\,{\sc iv}, Si\,{\sc iii}, O\,{\sc vi}, O\,{\sc vii} and O\,{\sc viii}) and their imprints on synthetic sky maps constructed from the reference frames commonly used by observers. Our results indicate the presence of such a dipole for gas outside the Milky Way halo, favouring a paradigm in which the Milky Way is moving against the gas in the direction of the barycentre, while moving away from it in the opposite direction. This pattern is clearer for the higher oxygen ions, which preferentially trace hot gas. On the other hand, we observe a slight asymmetry in the pressure profiles in both directions, indicating higher pressures in the inner regions of the Local Group.}


\begin{document}

\maketitle

\section{Introduction}

The Local Group (LG) is the name given to our cosmological vicinity, which includes our own galaxy, the Milky Way (MW) and the slightly more massive neighbour Andromeda (M31), as well as several smaller galaxies. The MW and M31 follow a collision course with a relative radial velocity of $-109 \pm 4.4\,$km\,s$^{-1}$ \citep{vanderMarel12} and are expected to undergo their first core passage in the next few Gyr \citep{Salomon21}.

LG kinematics are expected to be dominated by the radial motion of the MW and M31, and as such they should be more evident in the preferred direction joining both galaxies. From the MW, M31 lies in the general barycentre direction, and observations show a velocity dipole displayed by galaxies in the general barycentre direction and its antipode in the sky (which we will call the antibarycentre direction), which is understood as a generalized flow of galaxies towards the LG's barycentre. \cite{Richter17} proposed a scenario in which the intragroup medium (IGrM) gas follows this global motion towards the barycentre of the LG, supported by observations from the Cosmic Origins Spectrograph (COS) installed onboard the Hubble Space Telescope (HST) as this velocity dipole can also be seen for H\,{\sc i} absorbers in high velocity clouds (HVCs). Further evidence provided by \cite{Bouma19} suggests that some of the absorbers responsible for this dipole can be associated with gas beyond the MW's circumgalactic medium (CGM).

In previous work \citep{Biaus22}, we examined the plausibility of this scenario in three of the {\sc Hestia} simulations, finding agreement with observed trends. In this work we focus on one of the realizations (the one labeled $17\_11$), examining the role of six distinct ionic species in shaping the dipole pattern to see if it can be associated with IGrM gas. We do this by evaluating the spatial distribution of these ions within the simulations (e.g., whether they reside inside or beyond MW-like galactic halos) while also considering their kinematic properties.

\section{{\sc Hestia} simulations}

\begin{figure}
    \centering
    \includegraphics[width=.9\columnwidth]{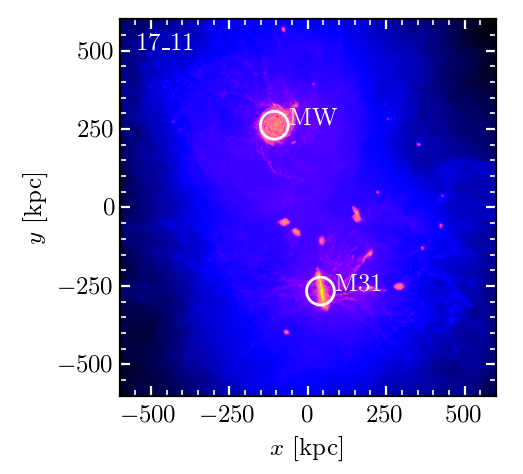}
    \caption{Projected gas density in the LG realisation $17\_11$ from the {\sc Hestia} high-resolution suite.}
    \label{fig:LG}
\end{figure}

{\sc Hestia} simulations were run using the cosmological moving-mesh code {\sc Arepo} \citep{Springel10, Weinberger20} and aim to resemble the LG. This is done by choosing initial conditions (ICs) such that the main observed cosmographic features (i.e. the local void, the local filament and the Virgo cluster) are present at $z=0$. Another criterion that the simulations must fulfill is that the MW and M31 are approaching each other at the present time. For this work, we focus on the high-resolution realisation, labeled $17\_11$, as it has the most accurate relative velocity for the MW and M31 ($v_{rad}$ = $-102 \,$km\,s$^{-1}$ at $z=0$). In Fig. \ref{fig:LG}, we show the projected gas density in this simulation. The MW and M31 visibly dominate the LG, but there are several smaller galaxies too. For more details on this simulation in particular, and the rest of the {\sc Hestia} project, we refer the reader to the original publication of \cite{Libeskind20}.

\section{Results}

\subsection{Ion kinematics relative to the LG’s barycentre}

\begin{figure*}[h!]
    \centering
    \includegraphics[width=2\columnwidth]{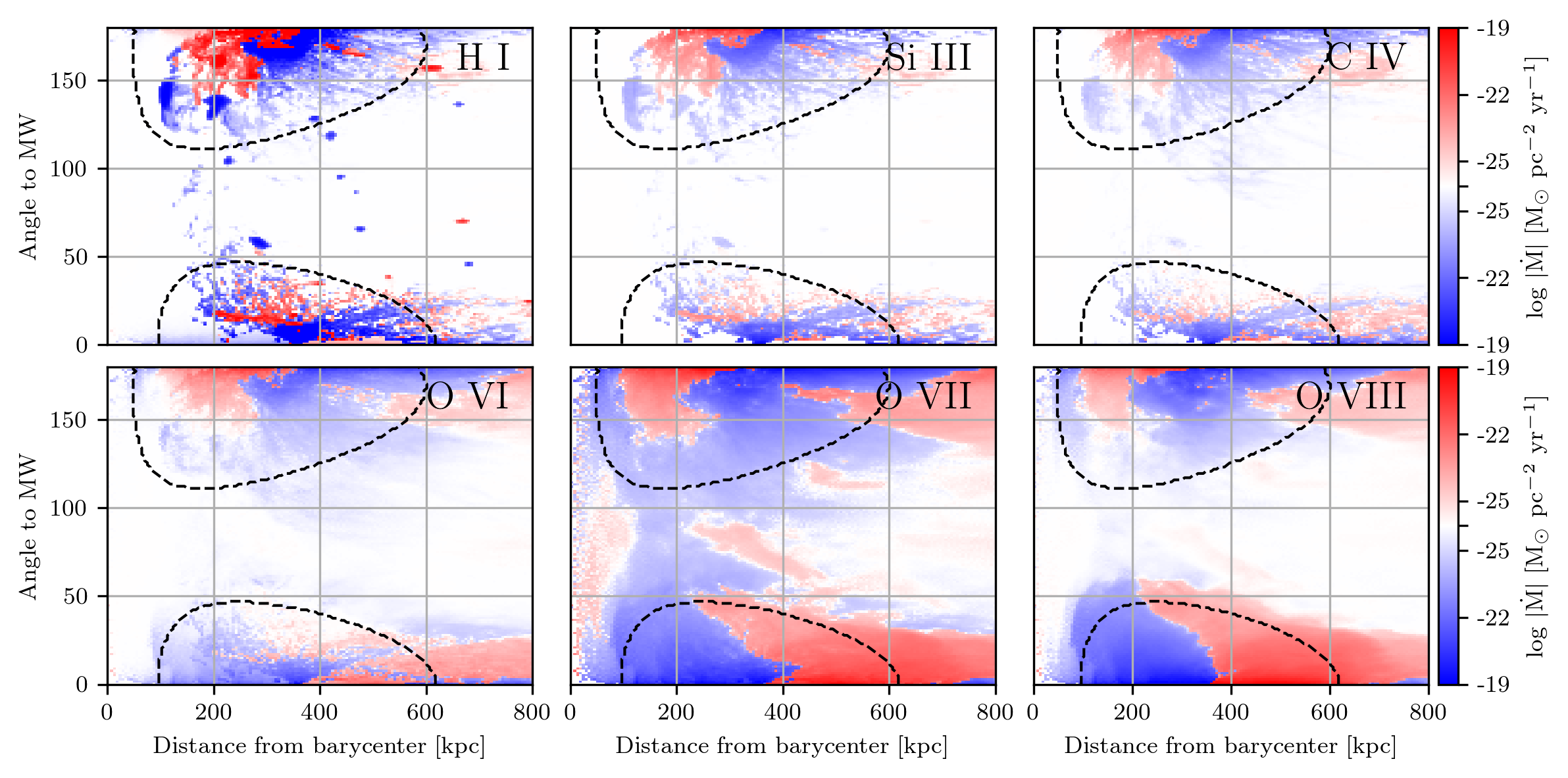}
    \caption{Radial mass flux as seen from the LG barycentre for each ion. Negative flux (blue) represents gas moving towards the LG barycentre and positive flux (red) represents material moving away from it. Note that the numbers in the colorbars represent the symmetric \emph{exponent} of the mass flux in the inflow and outflow cases. In all panels, black dashed lines enclose the CGM of the MW and M31 up to the corresponding $R_{200}$ values.
    }
    \label{fig:LG_flow}
\end{figure*}

To understand the global motion of the gas within the LG, we analyse the mass flux of the six different ions relative to the LG’s barycentre: H\,{\sc i}, Si\,{\sc iii}, C\,{\sc iv},  O\,{\sc vi}, O\,{\sc vii} and O\,{\sc viii}\footnote{The distribution of ions in the simulation may not be entirely realistic, as it can be influenced by subgrid galactic outflows, among other factors. The general trends should, however, be similar to the observed ones.}. In Fig. \ref{fig:LG_flow},
 we show the distance and angular distribution of radial mass inflow of each ion with respect to the LG barycentre. Here we measure angles for each gas cell relative to the MW's position from the barycentre, so the region enclosed by the dashed line centred at $0^{\circ}$ represents the MW's CGM and the one enclosed by the dashed line centred near $180^{\circ}$ represents the M31's CGM.

H\,{\sc i} kinematics is concentrated around galaxies, as is expected for an ion that traces cold, dense gas. Si\,{\sc iii} and C\,{\sc iv} exhibit similar kinematic patterns, with C\,{\sc iv} mass flux extending slightly further out beyond the main haloes than for Si\,{\sc iii}. Oxygen high ions, particularly O\,{\sc vii} and O\,{\sc viii}, trace the strong, hot-gas outflows present in the MW in this simulation, with O\,{\sc vii} being the ion which shows the highest mass flux beyond the two main haloes. O\,{\sc vii} also displays significant mass flux towards the LG barycentre in the directions perpendicular to the MW-M31 line, so we suggest it might be the most suitable ion to study IGrM kinematics.

\subsection{Placing an observer in the simulation}

To mimic observations, we place an observer in the simulated MW. This allows us to compare the simulation kinematics with the observed trends regarding absorbers in the general barycentre-antibarycentre direction. Here, we define the position of the Sun in the simulation so that M31 has the same galactic longitude as the observed one and the Sun is at a distance of $8$~kpc from the centre of the simulated MW. 

\begin{figure*}
    \centering
    \includegraphics[width=2\columnwidth]{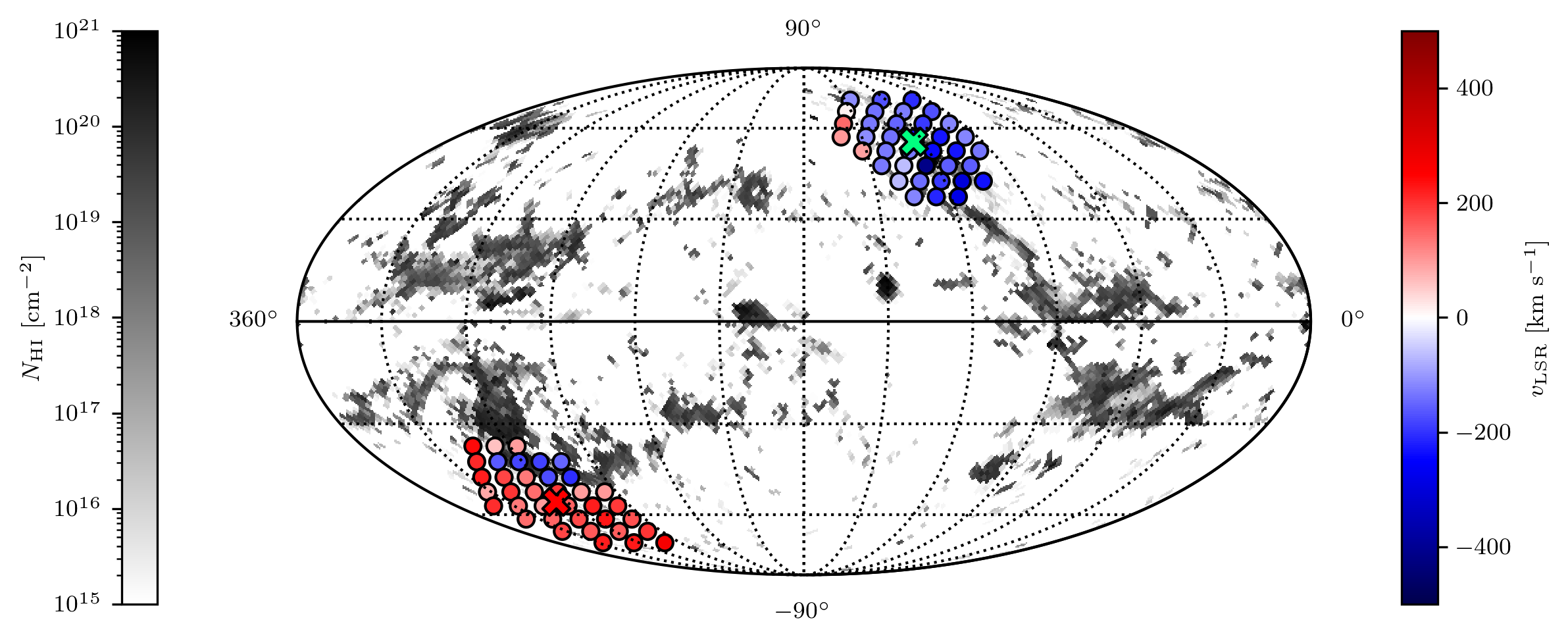}\caption{
    Column density map of H\,{\sc i} in simulation $17\_11$ after filtering out the gaseous disc, as seen by a simulated observer from Sun's position. Colour-coded LSR gas velocities in sightlines evenly spread towards the general barycentre (green cross) and antibarycentre (red cross) directions are also shown.
    }
    \label{fig:fig1}
\end{figure*}

From this position, we project lines towards the barycentre and in the opposite direction (antibarycentre) to evaluate the properties of the potential absorbers in these directions within the simulation. To filter out the disc material, we mimic the method described in \cite{Westmeier18}, which involves excluding, along each line of sight, gas with velocities consistent with a cylindrical rotation model for the disc. In Fig. \ref{fig:fig1}, we show a synthetic column density map of H\,{\sc i} after the disc material has been filtered out. Overplotted in the direction of the barycentre (antibarycentre), marked by a green (red) cross, gas velocity seen from the Local Standard of Rest\footnote{To define the Local Standard of Rest, we asume that the Sun is travelling with purely tangential velocity at a speed given by the simulated galaxy's rotation curve at $8$~kpc.} ($v_{\rm LSR}$) is shown. As it can be seen in the figure, the observed $v_{\rm LSR}$ dipole is replicated in this simulation and, although Galactic rotation contributes to this pattern, it doesn’t fully account for it (neither in observations nor in the simulations). This dipole is in part due to the motion of the MW across the LG. 

\subsection{Pressure profile and $v_{GSR}$}

\begin{figure}
    \centering
    \includegraphics[width=\columnwidth]{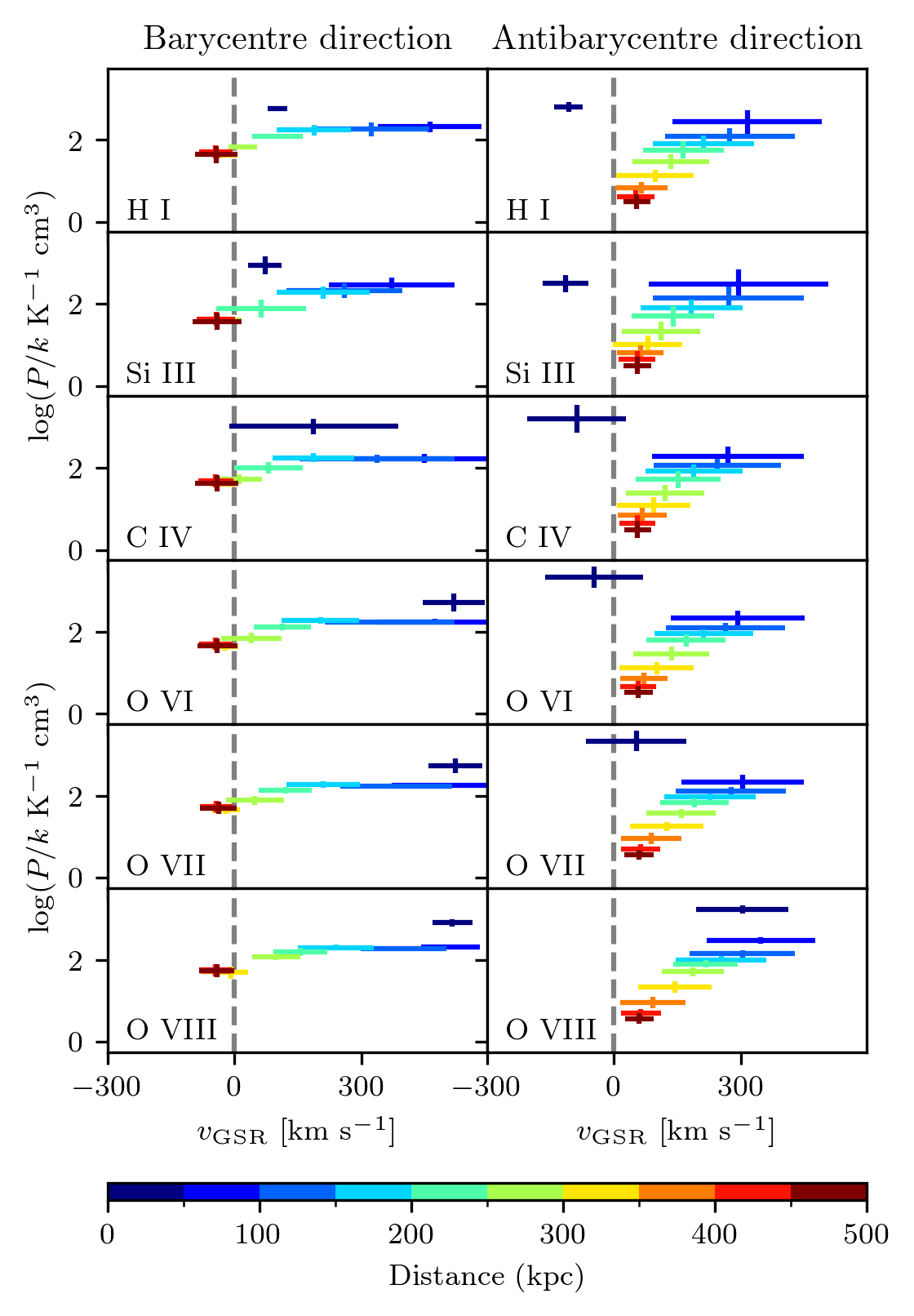}
    \caption{Distance-binned pressure and average line-of-sight velocites in the barycentre and antibarycentre directions for the GSR in realisation $17\_11$.}
    \label{fig:pressure_crosses}
\end{figure}

In \cite{Bouma19}, using HST/COS survey data, the authors derived pressure and density values for a subset of absorbers in the general barycentre-antibarycentre direction which contribute to the $v_{\rm LSR}$ dipole, indicating that some of the absorbers in the barycentre direction might be located beyond the MW's virial radius. This result would support the scenario proposed by \cite{Richter17}, where the observed dipole is related to the MW motion through the LG.

In Fig. \ref{fig:pressure_crosses}, we show the average pressure and velocity values for lines-of-sight in the barycentre and antibarycentre directions, weighted by the mass of each species, in radial bins of $50\,$~kpc from Sun's location for LG realisation $17\_11$. Expectedly, pressure decreases with distance for all six ions, but it does so in an anisotropic way beyond $r_{200} \approx 250$~kpc: pressure values are systematically lower in the antibarycentre direction as we are probing gas that lies towards the outskirts of the LG. Notice that in \cite{Bouma19} potential IGrM absorbers were found in the barycentre direction (those with low enough pressure values to be located beyond the MW's $R_{200}$), while this simulation suggests that lower pressure values should be expected in the opposite direction corresponding to the antibarycentre one. This suggests that absorbers in the antibarycentre direction aren't as easy to detect observationally and that further data would be necessary to assess the overall LG kinematics.

In regards to the velocity of the gas as seen from the Galactic Standard of Rest (GSR), we find that, beyond the halo, the gas shows $v_{\rm GSR}$ with negative (positive) values in the barycentre (antibarycentre) direction. This supports the scenario proposed in \cite{Richter17} in which the MW rams into gas at the LG barycentre (leading to negative relative velocities), while leaving behind the gas towards the outskirts of the LG  (leading to positive relative velocities). 

The most noticeable kinematic differences among the different ions are observed in the innermost regions of the galaxy, represented by the darkest blue crosses in Fig. \ref{fig:pressure_crosses}. Cold gas tracing ions, such as H\,{\sc i}, Si\,{\sc iii}, and C\,{\sc iv}, exhibit low $v_{\rm GSR}$ values within distances up to 50~kpc. In contrast, hot gas tracing high ions, i.e. O\,{\sc vi}, O\,{\sc vii}, and O\,{\sc viii}, display higher $v_{\rm GSR}$ values, especially in the direction of the barycentre. In particular, these higher $v_{\rm GSR}$ values are associated with strong MW outflows in this realisation, comprised of very hot gas. Differences in the $v_{\rm GSR}$ values between left and right panels within the halo evidence that these outflows are not isotropic.

\section{Discussion}

In this study, we analysed the kinematic properties of six distinct ionic species: H\,{\sc i}, Si\,{\sc iii}, and C\,{\sc iv} (tracers of cold gas), as well as O\,{\sc vi}, O\,{\sc vii}, and O\,{\sc viii} (tracers of hot gas). Our research was based on the $17\_11$ high-resolution simulation from the {\sc Hestia} project. This specific simulation was selected due to its comparable relative radial velocity between the two main galaxies, aligning with the observed velocity between the MW and M31. 

Firstly, we found that the MW's CGM strongly influences the observed kinematic patterns. Column densities for cold gas-tracing ions are notably high within the CGM, whereas galactic outflows exhibit much higher velocities compared to predictions for IGrM gas. This fact clearly complicates efforts to probe the gas distribution beyond the MW's CGM, unless thermodynamical conditions for absorbers can accurately be derived, as done in studies like \cite{Bouma19}.

Cold gas-tracing ions (such as H\,{\sc i}, Si\,{\sc iii}, and C\,{\sc iv}) display kinematics predominantly shaped by the disc and CGM gas. The mass flux of these ions, as observed from the LG's barycentre, is primarily concentrated within the CGM of the candidate galaxies. As a result, detecting IGrM signals in these ions may prove challenging. In contrast, hot gas-tracing ions, including O\,{\sc vi}, O\,{\sc vii}, and O\,{\sc viii}, exhibit kinematics that extend into the intergalactic medium and IGrM. Specifically, in the simulated MW, these ions trace CGM outflows, with O\,{\sc vii} showing significant mass flux beyond the CGM of both the MW and M31.

Lastly, our results support a scenario where the MW rams into gas near the LG's barycentre while simultaneously receding from gas at its outskirts. This scenario aligns with the observed $v_{\rm LSR}$ dipole in the barycentre-antibarycentre direction. We also found that, in this particular realisation, pressure values are systematically lower in the antibarycentre direction, which may complicate the interpretation of observations as there is a lack of potential IGrM absorbers in this direction. Consequently, a larger catalogue of absorbers will be required to better map the kinematics of the IGrM.

\begin{acknowledgement}
S.E.N. and C.S. are members of the Carrera del Investigador Cient\'{\i}fico of CONICET. They acknowledge funding from Agencia Nacional de Promoci\'on Cient\'{\i}fica y Tecnol\'ogica (PICT-2021-GRF-TI-00290). S.E.N also acknowledges support from CONICET through grant PIBAA R73734.
\end{acknowledgement}

\bibliographystyle{baaa}
\small
\bibliography{bibliografia}

\begin{thebibliography}{9}
\providecommand{\natexlab}[1]{#1}

\bibitem[{{Biaus} et~al.(2022)}]{Biaus22}
{Biaus} L., et~al., 2022, \mnras, 517, 6170

\bibitem[{{Bouma} et~al.(2019){Bouma}, {Richter} \& {Fechner}}]{Bouma19}
{Bouma} S.J.D., {Richter} P., {Fechner} C., 2019, \aap, 627, A20

\bibitem[{{Libeskind} et~al.(2020)}]{Libeskind20}
{Libeskind} N.I., et~al., 2020, \mnras, 498, 2968

\bibitem[{{Richter} et~al.(2017)}]{Richter17}
{Richter} P., et~al., 2017, \aap, 607, A48

\bibitem[{{Salomon} et~al.(2021)}]{Salomon21}
{Salomon} J.B., et~al., 2021, \mnras, 507, 2592

\bibitem[{{Springel}(2010)}]{Springel10}
{Springel} V., 2010, \mnras, 401, 791

\bibitem[{{van der Marel} et~al.(2012)}]{vanderMarel12}
{van der Marel} R.P., et~al., 2012, \apj, 753, 8

\bibitem[{{Weinberger} et~al.(2020){Weinberger}, {Springel} \& {Pakmor}}]{Weinberger20}
{Weinberger} R., {Springel} V., {Pakmor} R., 2020, \apjs, 248, 32

\bibitem[{Westmeier(2018)}]{Westmeier18}
Westmeier T., 2018, MNRAS, 474, 289

\end{thebibliography}
 
\end{document}